\documentclass[onecolumn,aps,prd,amssymb,eqsecnum,nofootinbib,superscriptaddress,showpacs,floatfix]{revtex4}
\pdfoutput=1
\newlength{\picwidth}
\usepackage{epsfig}
\usepackage{amsmath}
\usepackage{dcolumn}
\usepackage{amssymb}
\usepackage{amsfonts}
\usepackage{enumerate}
\usepackage{bm}
\makeatletter
\begin{document}
\def\be{\begin{equation}}
\def\ee{\end{equation}}
\def\ba{\begin{eqnarray}}
\def\ea{\end{eqnarray}}

\title{Second-order weak lensing from modified gravity}

\author{R. Ali Vanderveld}
\affiliation{Jet Propulsion Laboratory, California Institute of Technology, Pasadena, CA 91109}
\affiliation{California Institute of Technology, 1200 East California Boulevard, Pasadena, CA 91125}
\affiliation{Kavli Institute for Cosmological Physics, Enrico Fermi Institute, University of Chicago, Chicago, IL 60637}
\author{Robert R. Caldwell}
\affiliation{Department of Physics and Astronomy, Dartmouth College, Hanover, NH 03755}
\author{Jason Rhodes}
\affiliation{Jet Propulsion Laboratory, California Institute of Technology, Pasadena, CA 91109}
\affiliation{California Institute of Technology, 1200 East California Boulevard, Pasadena, CA 91125}

\begin{abstract}

We explore the sensitivity of weak gravitational lensing to second-order corrections to the spacetime metric within a cosmological adaptation of the parameterized post-Newtonian framework. Whereas one might expect nonlinearities of the gravitational field to introduce non-Gaussianity into the statistics of the lensing convergence field, we show that such corrections are actually always small within a broad class of scalar-tensor theories of gravity. We show this by first computing the weak lensing convergence within our parameterized framework to second order in the gravitational potential, and then computing the relevant post-Newtonian parameters for scalar-tensor gravity theories. In doing so we show that this potential systematic factor is generically negligible, thus clearing the way for weak lensing to provide a direct tracer of mass on cosmological scales for a wide class of gravity theories despite uncertainties in the precise nature of the departures from general relativity.

\end{abstract}


\maketitle

\section{Introduction}

Weak gravitational lensing, whereby galaxy images are altered due to the gravitational influence of the mass along the line of sight, is a powerful probe of the ``dark sector" of cosmology, with promising results in recent years, e.g.~\cite{Fu:2007qq,Kilbinger:2008gk,Massey:2007gh,Massey:2007wb,Schrabback:2009ba}. (See Refs.~\cite{Takada:2003ef,Hoekstra:2008db,Massey:2010hh} for recent reviews.) In general relativity (GR), the weak lensing distortion field provides a direct tracer of the underlying matter distribution. However, modifications to gravity theory can conceivably alter the way that mass curves spacetime and thus the way that null geodesics correspond to the matter distribution. The linear order effect, often referred to as the ``gravitational slip,"  has already been explored in the literature \cite{Daniel:2010ky}, with weak lensing and other datasets. But what happens at higher order? Higher-order lensing statistics are useful for breaking degeneracies and probing primordial non-Gaussianity (e.g. \cite{Takada:2003ef,Berge:2009xj}), but could modifications to GR complicate such efforts by further altering the way that lensing traces mass? On the other hand, could we conceivably use higher-order weak lensing statistics to probe nonlinear gravitational dynamics?

We aim to answer these questions by utilizing a cosmological adaptation of the parameterized post-Newtonian (PPN) framework \cite{Will:1993ns,Will:2001mx}. For example, consider the following special case for the static weak-field metric:
\ba
ds^2&=&g_{\mu\nu}dx^{\mu}dx^{\nu}\nonumber\\
&=&-\left(1 - \frac{2}{c^2}U + \frac{2 \beta}{c^4}U^2\right) c^2 dt^2 + \left(1 + \frac{2 \gamma}{c^2}U + \frac{3 \varepsilon}{2 c^4} U^2\right)d \vec x^2~,
\label{eqn:PPNmetric}
\ea
where $U$ is the usual Newtonian gravitational potential and the powers of $1/c^2$ label the order of the perturbation from Minkowski spacetime. The parameters take on the values $\gamma = \beta = \varepsilon = 1$ in the case of GR.  At second order, $\beta$ and $\varepsilon$ represent the first nonlinear effects of gravity\footnote{Because non-relativistic objects move with $|d\vec x| \ll c dt$, the $\varepsilon$ term enters at a higher order than does the $\beta$ term. For this reason the $\varepsilon$ term is often neglected in investigations of alternative theories of gravity. However, the $\varepsilon$ term is necessary for a self-consistent analysis of the equations of motion of light, i.e. gravitational lensing.}. In cosmology we would like to determine if new laws of gravitation are responsible for cosmic acceleration, possibly replacing the need for dark energy \cite{Carroll:2003wy,Caldwell:2009ix}. If new gravitational phenomena are at play, then we should expect a departure from GR in the cosmological analogues of $\gamma, \beta$, and $\varepsilon$, which may be manifest in the patterns of gravitational lensing of light from galaxies, clusters, and the cosmic microwave background (CMB).

The nonlinearity of the gravitational field introduced by the $\beta$ and $\varepsilon$ terms translates into a non-Gaussianity of the statistics of cosmological fluctuations. Indeed, standard single-field inflation models predict nearly Gaussian initial cosmological fluctuations \cite{Creminelli:2004yq}.  As such, probes of non-Gaussianity provide a crucial test of the inflationary paradigm, and thus much effort has been put into constraining non-Gaussianity in both the early- and late-time universe. Local type non-Gaussianity is typically parameterized by the parameter $f_{\rm NL}$, where on sub-horizon scales the Newtonian potential $\phi$ (equal to $U$ above and below) can be re-written in terms of a Gaussian field $\phi_0$:
\be
\phi(\vec{x})=\phi_0(\vec{x})+f_{\rm NL}\left[\phi_0(\vec{x})^2-\left\langle \phi_0^2(\vec{x}) \right\rangle\right]~.
\label{fnl}
\ee
If we take the GR metric and replace $U \to U + \Delta f_{\rm NL} U^2$, then we find
\be
ds^2 = -\left[1 - \frac{2}{c^2}U + \frac{2}{c^2}(1 - \Delta f_{\rm NL})U^2 \right] c^2 dt^2 + \left[1 + \frac{2}{c^2}U + \frac{2}{c^4}\left(\frac{3}{4} + \Delta f_{\rm NL}\right)U^2\right] d\vec x^2
\label{eqn:fnl1}
\ee
to second order in $U$. Comparing this with the line element (\ref{eqn:PPNmetric}), we may infer that additional non-Gaussianity beyond GR is introduced by the $\beta$ and $\varepsilon$ terms. Such additional non-Gaussianity could conceivably be confused with that of primordial origin. In this paper we determine how gravitational lensing can probe this effect; by computing $\beta$ and $\varepsilon$ for a wide class of scalar-tensor gravity theories, we further show that such corrections to $f_{\rm NL}$ are generically negligible for the majority of the modified gravity literature. In doing so, we further show how the second-order lensing convergence depends on the ``gravitational slip" $\gamma$. 

The organization and results of this paper are as follows. In Section II, we compute the weak lensing convergence to second order in the gravitational potential $U$, or to ${\cal O}$(${1/c^4}$), within a cosmological extension of the PPN formalism so that we may find how nonlinear lensing and non-Gaussianity depend on modifications of GR.  Our first main result is our final answer for the weak lensing convergence, which is is given in Eqs.~(\ref{kappatotal}-\ref{kappaGM}).  Note that throughout this paper $\gamma$ will refer to the PPN parameter and not to the cosmic shear field. Then in Section III we study a wide class of scalar-tensor theories of gravity, wherein we provide a derivation of the relevant post-GR parameters and show that the combination of parameters important for lensing negligibly differs from the GR expectation.  We conclude in Section IV with a discussion of our results.

\section{Nonlinear gravitational lensing}

Generalizing the PPN metric (\ref{eqn:PPNmetric}) to allow for non-static gravitational potentials, we write
\be
ds^2=-\left(1-\frac{2}{c^2}U+\frac{2\beta}{c^4} U^2+\frac{\beta'}{c^4}\Phi_2 \right)c^2 dt^2+\frac{2}{c^3} V_i \,c dt \,dx^i+
\left(1+\frac{2\gamma}{c^2} U+\frac{3\varepsilon}{2c^4} U^2+\frac{\varepsilon'}{c^4}\Phi_2\right) {d\vec x}^2~,
\label{eqn:genmetric}
\ee
to order ${\cal O}(1/c^4)$ \cite{Will:1993ns,Will:2001mx}, where the potentials are defined such that GR is recovered when $\gamma=\beta=\varepsilon=\beta'=\varepsilon'=1$, the necessary PPN parameters are absorbed into the definition of the gravitomagnetic potential $V_i$, and $\Phi_2$ is the post-Newtonian scalar potential as in \cite{Will:2001mx}. To start, we assume that the background spacetime is Minkowski, but we will generalize to the case of an expanding Universe at the end of the calculation.  We also now set $G=c=1$, but leave the powers of $1/c$ in place for post-Newtonian bookkeeping purposes.

\begin{figure}[h]
\includegraphics[scale=0.4]{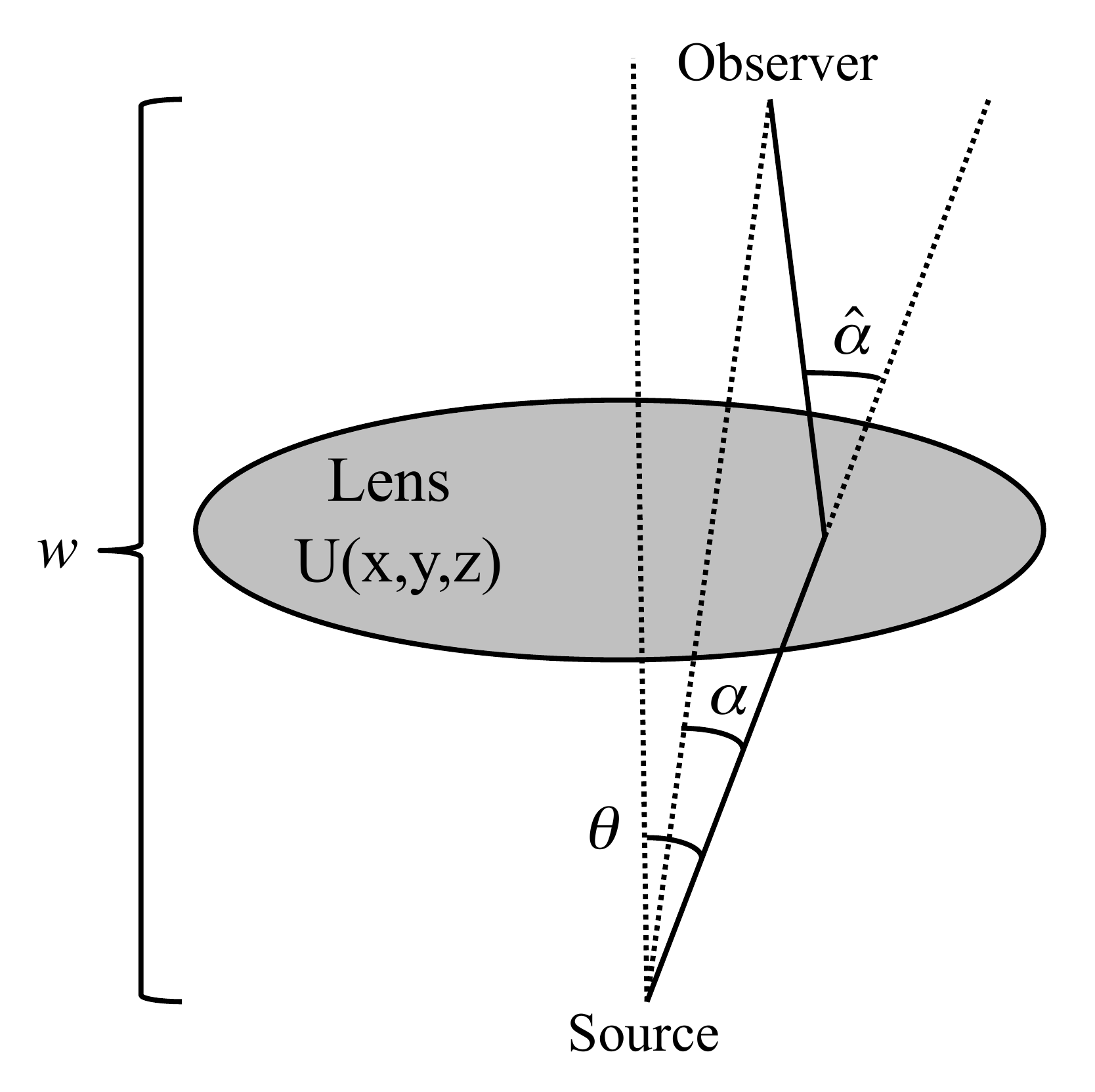}
\caption{A sketch of our setup.  The observer is a comoving distance $w$ away from the source, $\theta$ is the angular coordinate on the sky, $\hat{\alpha}$ is the deflection angle felt by the ray, and $\alpha$ is the angular change in sky position as seen by the observer.}
\label{fig:sketch}
\end{figure}

The propagation of photons is determined by the null geodesic equation
\be
\frac{dk^{\alpha}}{d\lambda}=-\Gamma_{\mu\nu}^{\alpha}k^{\mu}k^{\nu}~,
\label{eqn:null}
\ee
where $k^{\alpha}(\lambda)=dx^{\alpha}/d\lambda$ is the photon 4-momentum, $x^{\alpha}(\lambda)$ is the light ray trajectory, $\lambda$ is an affine parameter, and the connection is for the full metric.  Let us assume that the unperturbed ray moves along the positive x-axis,  and we will place the source at $x=0$ and the observer at some location $x$, corresponding to a comoving distance $w(x)$.  We enforce the null geodesic condition, $g_{\mu\nu}k^{\mu}k^{\nu}=0$, and normalize $k^\mu = (k^t,k^x,k^y,k^z)=(1,1,0,0)$ for the unperturbed ray.  The geometry of the generic lensing scenario is sketched in Figure~\ref{fig:sketch}.  The angle $\hat{\alpha}$ is the deflection angle felt by the ray and $\alpha$ is the angular change in sky position as seen by the observer.

We can change the independent variable from $\lambda$ to $x$ using the following relation:
\be
\frac{d}{d\lambda}=\frac{dx}{d\lambda}\frac{d}{dx}=k^x\frac{d}{dx}~.
\label{eqn:change}
\ee
Thus, if we define
\be
W^{\alpha}(\vec{x})=-\frac{\Gamma^{\alpha}_{\mu\nu}k^{\mu}k^{\nu}}{k^{x}}~,
\label{eqn:Weqn}
\ee
then we can integrate Eq.~(\ref{eqn:null}) along the ray trajectory, with the help of Eqs.~(\ref{eqn:change}) and (\ref{eqn:Weqn}), to find the 4-momentum as a function of the independent variable $x$,
\be
k^{\alpha}(x)=\int^x_{0}dx'W^{\alpha}\left[\vec{x}\left(x'\right)\right]~ .
\ee
Here we are integrating along the perturbed ray trajectory $\vec{x}(x)$,
\be
x^{\alpha}(x)=\int^x_{0}dx'\frac{k^{\alpha}\left[\vec{x}\left(x'\right)\right]}{k^x\left[\vec{x}\left(x'\right)\right]}~.
\ee
which we are also parameterizing as a function of the x-coordinate along the ray.  Within the perturbative post-Newtonian framework, these equations can be solved iteratively to get the full ray trajectory $\vec{x}(x)$ to any order in $1/c$.  Once we know the ray position at the observer, $x^i(x)$, then the angle $\alpha^i$ (see Figure~\ref{fig:sketch}) is such that $x^i(x)=x\alpha^i(x)$, where now $i=y,z$ if we use the flat sky approximation.  Thus
\be
\alpha^i(x)=\frac{1}{x}\int^x_{0}\frac{dx'}{k^x\left[\vec{x}\left(x'\right)\right]}\int^{x'}_{0}dx''W^i\left[\vec{x}\left(x''\right)\right]
\ee
is the deflection angle seen by the observer.

\subsection{The deflection angle}

Generalizing to the cosmological scenario, whereupon the metric is replaced by $g_{\mu\nu} \to a^2(t) g_{\mu\nu}$ so that all times are conformal and coordinate positions are comoving, we now denote $w$ as the comoving distance to the image source and $\vec\theta$ as the location on the observer's image plane.  Then we find
\be
\alpha_i(\vec\theta,w) = \frac{1}{c^2}(1+\gamma) \int_0^w dw' \left(\frac{w-w'}{w}\right) \partial_i U[w'\vec\theta,w'] + {\rm 2PN~terms}+\ldots
\label{eqn:1PNlens}
\ee
to first post-Newtonian (1PN) order.  This is the standard result \cite{Bartelmann:1999yn}.  Note that we have pulled $\gamma$ out of the integral, thus ignoring its possible spatial dependence. Generally $\gamma$ will take on different values on sub-galactic scales than on larger scales so as to maintain the GR limit in the well-tested very small-scale regime. However, such a small-scale transition can be safely ignored for the purposes of the weak lensing observations discussed here, which are limited in resolution to scales of $\sim 1$ arcminute, or 0.5 Mpc at $z=1$. The expression above also uses the Born approximation, whereby the photon moves along the unperturbed geodesic, $\vec x_{\rm Born} = w \vec\theta$.  In fact, the linearly-perturbed geodesic is $\vec x_{\rm Linear} = \vec x_{\rm Born} + \delta\vec x$ where $\delta x_i = -w \alpha_i$. To implement this correction, we Taylor expand the potential:
\ba
\alpha_i(\vec\theta,w) &=&  \int_0^w dw' \left(\frac{w-w'}{w}\right) \left( \frac{1}{c^2}(1+\gamma) \partial_i U[w'\vec\theta,w']
-  \frac{1}{c^4}(1+\gamma)^2 \partial_{ij}U[w'\vec\theta,w']\int_0^{w'} dw'' (w'-w'') \partial_jU[w''\vec\theta,w'']\right)\nonumber\\
& &+ {\rm 2PN~terms}+\ldots~.
\ea
Now letting the unit 3-vector $\vec{n}$ denote the direction of propagation of the unperturbed ray, the displacement vector is
\ba
\alpha_i(\vec\theta,w) &=& \int_0^w dw'\left(\frac{w-w'}{w}\right)\left[ \frac{1+\gamma}{c^2} \partial_i U[w'\vec\theta,w']-\frac{(1+\gamma)^2}{c^4} \partial_{ij}U[w'\vec\theta,w'] \int_0^{w'} dw'' (w'-w'') \partial_jU[w''\vec\theta,w''] \right] \cr\cr
&+&\int_0^w dw' \left(\frac{w-w'}{w}\right)\left[\frac{1}{c^3}\left(\vec{n}\times\left\{\vec{\nabla}\times\vec{V}[w'\vec\theta,w']\right\}\right)_i+\left(\frac{6-4\beta+3\varepsilon-6\gamma^2}{2 c^4}\right)U[w'\vec\theta,w']  \partial_i U[w'\vec\theta,w'] \right] \cr\cr
&+& \int_0^w dw' \left(\frac{w-w'}{w}\right)\frac{1}{c^4}\left[\frac{\varepsilon'-\beta'}{2}\partial_i\Phi_2[w'\vec\theta,w'] \right]- \int_0^w \left(\frac{dw'}{w}\right)\left[\frac{1-\gamma^2}{c^4}U[w'\vec\theta,w'] \int_0^{w'} dw''  \partial_i U[w''\vec\theta,w'']\right]\cr \cr
&-&\int_0^w dw' \left(\frac{w-w'}{w}\right)\frac{1}{c^4}\left[\dot{V}_i+2\gamma(1+\gamma)\vec{n}\cdot\vec{\nabla}U[w'\vec\theta,w']  \int_0^{w'} dw'' \partial_i U[w''\vec\theta,w'']  \right]
\label{2pn}
\ea
to 2PN order, where all of these potentials are evaluated along the unperturbed ray trajectory.  The first line contains the linear order answer plus the Born correction and the second line contains the gravitomagnetic term.  The first line is studied in Ref.~\cite{Krause:2009yr} and the gravitomagnetic term is studied in Ref.~\cite{Schaefer:2005up}.

The integrands are evaluated along the line of sight, and so some simplifications may be made. Note that
\be
\left(\vec{n}\cdot\vec{\nabla}\right)U=\dot{U}+\frac{\partial}{\partial w'}U~,
\ee
where the overdot denotes a derivative with respect to $\tau(w')=w-w'$.  Then the second order term on the fourth line of Eq.~(\ref{2pn}) can be rearranged thusly:
\ba
& &\int_0^w dw' \left(\frac{w-w'}{w}\right)\vec{n}\cdot\vec{\nabla}U[w'\vec\theta,w']  \int_0^{w'} dw'' \partial_i U[w''\vec\theta,w'']\nonumber\\
& &=\int_0^w dw' \left(\frac{w-w'}{w}\right) \left( \dot{U}[w'\vec\theta,w'] + \frac{\partial}{\partial w'}U[w'\vec\theta,w'] \right) \int_0^{w'} dw'' \partial_i U[w''\vec\theta,w'']\nonumber\\
& &=\int_0^w dw' \left(\frac{w-w'}{w}\right) \dot{U}[w'\vec\theta,w']  \int_0^{w'} dw'' \partial_i U[w''\vec\theta,w''] + \int_0^w dw' \left(\frac{1}{w}\right) U[w'\vec\theta,w']  \int_0^{w'} dw'' \partial_i U[w''\vec\theta,w''] \nonumber\\
& &~ - \int_0^w dw' \left(\frac{w-w'}{w}\right) U[w'\vec\theta,w'] \partial_i U[w''\vec\theta,w'']~.
\ea
Using this substitution, we find the full deflection angle to second post-Newtonian (2PN) order to be
\ba
\alpha_i(\vec\theta,w) &=& \int_0^w dw' \, \left(\frac{w-w'}{w}\right)\frac{1}{c^2}(1+\gamma) \partial_i U[w'\vec\theta,w']\nonumber\\
&-&\int_0^w dw' \, \left(\frac{w-w'}{w}\right)\frac{1}{c^4}(1+\gamma)^2 \partial_{ij}U[w'\vec\theta,w']\int_0^{w'} dw'' \,(w'-w'')\partial_jU[w''\vec\theta,w'']  \nonumber\\
&+&\int_0^w dw' \, \left(\frac{w-w'}{w}\right)\left[ \frac{1}{2 c^4}(6-4\beta+4\gamma+3\varepsilon-2\gamma^2) U[w'\vec\theta,w']  \partial_i U[w'\vec\theta,w'] \right] \nonumber\\
&-&\int_0^w dw' \, \left(\frac{w-w'}{w}\right)\left[\frac{2}{c^4}\gamma(1+\gamma)\dot{U}[w'\vec\theta,w']  \int_0^{w'} dw'' \, \partial_i U[w''\vec\theta,w'']  \right] \nonumber\\
&-&\int_0^w dw'\, \left(\frac{1}{w}\right)\left[\frac{1}{c^4}(1+2\gamma+\gamma^2) U[w'\vec\theta,w'] \int_0^{w'} dw'' \, \partial_i U[w''\vec\theta,w'']\right]  \nonumber\\
&+&\int_0^w dw' \, \left(\frac{w-w'}{w}\right) \frac{1}{c^3}\left\{\vec{n}\times\left(\vec{\nabla}\times\vec{V}[w'\vec\theta,w']\right)\right\}_i \nonumber\\
&+&\int_0^w dw' \, \left(\frac{w-w'}{w}\right)\frac{1}{2 c^4}(\varepsilon'-\beta') \partial_i \Phi_2[w'\vec\theta,w'] \nonumber\\
&-&\int_0^w dw' \, \left(\frac{w-w'}{w}\right) \frac{1}{c^4}\dot{V}_i[w'\vec\theta,w']~.
\label{2pn2}
\ea

\subsection{An aside: Point mass case}

We can specialize the above results to the well-studied case of a static point mass.  A point mass located a distance $b$ away from the unperturbed ray trajectory produces a gravitational potential
\be
U(x,y,z)=\frac{m}{\sqrt{(x-x_l)^2+(y-b)^2+z^2}}
\ee
where we have arbitrarily placed the mass at $(x,y,z)=(x_l,b,0)$, meaning that there will be deflection in the y-direction only.  For the point mass, we also have $\Phi_2=0$ because it is static. We further assume that the source and observer are very far away from the lensing mass, meaning that we can safely take the lens position to be such that $x_l\gg 0$ and the observer position to be such that $x\rightarrow\infty$.  Plugging this all into the procedure outlined above simplifies the equations greatly, and we find the deflection angle to be
\be
\hat{\alpha} = (1+\gamma)\frac{2m}{c^2b} + \left(2+2\gamma-\beta+\frac{3}{4}\varepsilon\right)\frac{\pi m^2}{c^4b^2}
\ee
to 2PN order, where we have now calculated $\hat{\alpha}$ instead of $\alpha$ so as to compare directly with the literature.  The 1PN piece of this answer is the widely accepted result, and the 2PN piece agrees with Ref.~\cite{Epstein:1980dw}.

\subsection{Order of magnitude scalings}
\label{sec:oom}

A close inspection of the deflection angle formula above, Eq.~(\ref{2pn2}), permits us to estimate the scalings of the various terms as in Ref.~\cite{Dodelson:2005zj} and determine the relative contributions.  Note that $w\sim r_H$, where $r_H\sim3000h^{-1}~{\rm Mpc}$ is the Hubble radius, and spatial derivatives in the plane of the sky introduce factors of $1/\lambda$, where $\lambda$ is the mode wavelength.  For the perturbations which contribute the most to deflections, $\lambda\sim\lambda_{\rm max}\sim 30h^{-1}~{\rm Mpc}\sim r_H/100$.  
This also justifies our discard of time-derivative terms, under the implicit assumption that the PPN-like parameters evolve on a time scale comparable to the Hubble time. 
Also note that the rms amplitude of the gravitational potential on these scales is $U_{\rm rms}\sim 10^{-5}$.  With this knowledge in hand, Eq.~(\ref{2pn2}) contains the following terms:

\begin{itemize}

\item Line 1: $\int_0^w dw' \, \left(\frac{w-w'}{w}\right)\frac{1}{c^2}(1+\gamma) \partial_i U[w'\vec\theta,w']$

\noindent This is the leading order solution, which scales like $(r_H/\lambda)U$ and is thus of order $10^{-3}$ for the dominant mode.

\item Line 2: $\int_0^w dw' \, \left(\frac{w-w'}{w}\right)\frac{1}{c^4}(1+\gamma)^2 \partial_{ij}U[w'\vec\theta,w']\int_0^{w'} dw'' \,(w'-w'')\partial_jU[w''\vec\theta,w'']$

\noindent This is the Born correction, which scales like $(r_H/\lambda)^3U^2$.  That means that it is of order $10^{-4}$ for the dominant mode, making it roughly $10\%$ of the leading order term for that mode.  However, this term is significantly suppressed on larger scales.

\item Line 3: $\int_0^w dw' \, \left(\frac{w-w'}{w}\right)\left[ \frac{1}{2 c^4}(6-4\beta+4\gamma+3\varepsilon-2\gamma^2) U[w'\vec\theta,w']  \partial_i U[w'\vec\theta,w'] \right]$

\noindent Modulo the coefficients $\gamma, \beta,$ and $\varepsilon$ then this term scales like $(r_H/\lambda)U^2$, which is of order $10^{-8}$ for the dominant mode. An alternative theory of gravity would have to produce PPN-like parameters of magnitude $10^4$ or greater in order for this term to be as important as the Born correction. On the other hand, if the PPN-like parameters are $\lesssim 10^2$ always, then this term may be difficult to measure except on very large scales \cite{Bernardeau:2009bm} where $w\sim\lambda$ and the Born correction is suppressed by comparison.

\item Line 4: $\int_0^w dw' \, \left(\frac{w-w'}{w}\right)\left[\frac{2}{c^4}\gamma(1+\gamma)\dot{U}[w'\vec\theta,w']  \int_0^{w'} dw'' \, \partial_i U[w''\vec\theta,w'']  \right]$

\noindent Noting that $\dot{U}\sim U/r_H$, this term scales like $(r_H/\lambda)U^2$, which is roughly of order $10^{-8}$ for the dominant mode and even smaller for larger-wavelength modes.

\item Line 5: $\int_0^w dw'\, \left(\frac{1}{w}\right)\left[\frac{1}{c^4}(1+2\gamma+\gamma^2) U[w'\vec\theta,w'] \int_0^{w'} dw'' \, \partial_i U[w''\vec\theta,w'']\right]$

\noindent This term also scales like $(r_H/\lambda)U^2$.

\item Line 6: $\int_0^w dw' \, \left(\frac{w-w'}{w}\right) \frac{1}{c^3}\left\{\vec{n}\times\left(\vec{\nabla}\times\vec{V}[w'\vec\theta,w']\right)\right\}_i$

\noindent This is the gravitomagnetic term.  Note that $\vec{\nabla}\cdot\vec{V}\sim\dot{U}$ in the matter equations, meaning that $|\vec{V}|\sim (\lambda/r_H)U$.  Therefore this term scales like $U\sim 10^{-5}$, making it $\sim 1\%$ the size of the leading order term and $10$ times smaller than the Born correction.

\item Line 7: $\int_0^w dw' \, \left(\frac{w-w'}{w}\right)\frac{1}{2 c^4}(\varepsilon'-\beta') \partial_i \Phi_2[w'\vec\theta,w']$

\noindent Noting that $\Phi_2\sim U^2$, this term scales like $(r_H/\lambda)U^2\sim 10^{-8}$, which is a factor of $10^{5}$ smaller than the leading order term.

\item Line 8: $\int_0^w dw' \, \left(\frac{w-w'}{w}\right) \frac{1}{c^4}\dot{V}_i[w'\vec\theta,w']$

\noindent Noting that $|\vec{\dot{V}}|\sim |\vec{V}|/r_H$, this term scales like $|\vec{V}|\sim (\lambda/r_H)U$.

\end{itemize}

\noindent Therefore we can safely neglect terms 4, 5, and 7.
We also neglect terms containing a time derivative of the PPN-like parameters, as arising from lines 4 and 8, which are suppressed by a factor $(\lambda/r_H)$.
This leads to the final answer for the deflection angle,
\be
\alpha_i(\vec\theta,w) = \alpha^{(1)}_i(\vec\theta,w) + \alpha^{(2)}_i(\vec\theta,w) + \alpha^{({\rm GM})}_i(\vec\theta,w)~,
\label{2pn3}
\ee
where
\be
\alpha^{(1)}_i(\vec\theta,w)\equiv \frac{(1+\gamma)}{c^2} \int_0^w dw' \, \left(\frac{w-w'}{w}\right) \partial_i U[w'\vec\theta,w']
\ee
is the piece that is first order in $U$,
\ba
\alpha^{(2)}_i(\vec\theta,w)&\equiv& \int_0^w dw' \, \left(\frac{w-w'}{w}\right)\Bigg[\frac{(6-4\beta+4\gamma+3\varepsilon-2\gamma^2)}{2 c^4} U[w'\vec\theta,w'] \partial_i U[w'\vec\theta,w']\nonumber\\
& &~~~~~~~~~~~~~~~~~~~~~~~~~~-\frac{(1+\gamma)^2}{c^4} \partial_{ij}U[w'\vec\theta,w']\int_0^{w'} dw'' \,(w'-w'')\partial_jU[w''\vec\theta,w'']\Bigg]
\ea
is the piece that is second order in $U$, and
\be
\alpha^{({\rm GM})}_i(\vec\theta,w)\equiv \int_0^w dw'\, \left(\frac{w-w'}{w}\right) \left[ \frac{1}{c^3}\left\{\vec{n}\times\left[\vec{\nabla}\times\vec{V}[w'\vec\theta,w']\right]\right\}_i- \frac{1}{c^4}\dot{V}_i[w'\vec\theta,w']\right]
\ee
is the piece that depends on the gravitomagnetic potential.

\subsection{The distortion tensor and convergence}

The distortion tensor is
\be
\psi_{ij}=\frac{\partial\alpha_i}{\partial\theta_j}=\psi^{(1)}_{ij}+\psi^{(2)}_{ij}+\psi^{(GM)}_{ij}
\ee
where we note that
\be
\frac{\partial}{\partial\theta_j}U[w'\vec\theta,w']=w'\partial_j U[w'\vec\theta,w']~.
\ee
Therefore the linear order piece is
\be
\psi^{(1)}_{ij}=\frac{(1+\gamma)}{c^2} \int_0^w dw' \, g(w,w') \, \partial_{ij} U[w'\vec\theta,w']~,
\label{linearpsi}
\ee
where we have defined the weighting function to be
\be
g(w,w')\equiv \frac{w'(w-w')}{w}~.
\ee
The second order piece of the distortion tensor is
\ba
\psi^{(2)}_{ij}&=& \frac{(6-4\beta+4\gamma+3\varepsilon-2\gamma^2)}{2 c^4}\int_0^w dw' g(w,w') \left[ \partial_iU[w'\vec\theta,w']\partial_jU[w'\vec\theta,w'] + U[w'\vec\theta,w']\partial_{ij}U[w'\vec\theta,w']\right]\nonumber\\
& &-\frac{(1+\gamma)^2}{c^4} \int_0^w dw' g(w,w') \int_0^{w'} dw'' \frac{g(w',w'')}{w''}\left[w'\partial_{ijk}U[w'\vec\theta,w']\partial_kU[w''\vec\theta,w'']+w''\partial_{ik}U[w'\vec\theta,w']\partial_{jk}U[w''\vec\theta,w'']\right]\nonumber\\
&~&
\ea
and the gravitomagnetic piece is
\be
\psi^{({\rm GM})}_{ij}=\int_0^w dw' g(w,w') \left[ \frac{1}{c^3}\left\{\vec{n}\times\left[\vec{\nabla}\times\vec{V}[w'\vec\theta,w']\right]\right\}_{i,j} - \frac{1}{c^4}\dot{V}_{i,j}[w'\vec\theta,w']\right]~.
\ee

The lensing convergence of the light rays from the redshift slice corresponding to the distance $w$ is
\be
\kappa(\vec\theta)=\frac{1}{2}\psi_{ii}=\kappa^{(1)}(\vec\theta)+\kappa^{(2)}(\vec\theta)+\kappa^{({\rm GM})}(\vec\theta)
\label{kappatotal}
\ee
containing the following pieces:
\be
\kappa^{(1)}(\vec\theta) = \frac{(1+\gamma)}{c^2} \int_0^w dw' \, g(w,w') \, \nabla^2 U[w'\vec\theta,w']
\label{kappa1}
\ee
at first order in $U$,
\ba
\kappa^{(2)}(\vec\theta) &=& \frac{(6-4\beta+4\gamma+3\varepsilon-2\gamma^2)}{2 c^4}\int_0^w dw' g(w,w') \left\{ \left[\nabla U[w'\vec\theta,w']\right]^2 + U[w'\vec\theta,w']\nabla^2 U[w'\vec\theta,w']\right\}\nonumber\\
& &-\frac{(1+\gamma)^2}{c^4} \int_0^w dw' g(w,w') \int_0^{w'} dw'' \frac{g(w',w'')}{w''}\left[w'\partial_{iik}U[w'\vec\theta,w']\partial_kU[w''\vec\theta,w'']+w''\partial_{ik}U[w'\vec\theta,w']\partial_{ik}U[w''\vec\theta,w'']\right]~,\nonumber\\
&~&
\label{kappa2}
\ea
at second order in $U$, and
\be
\kappa^{({\rm GM})}(\vec\theta) = \int_0^w dw' g(w,w') \vec{\nabla}\cdot\left[ \frac{1}{c^3}\vec{n}\times\left[\vec{\nabla}\times\vec{V}[w'\vec\theta,w']\right] - \frac{1}{c^4}\dot{\vec{V}}[w'\vec\theta,w']\right]
\label{kappaGM}
\ee
to first order in the gravitomagnetic potential. Eqs.~(\ref{kappatotal}-\ref{kappaGM}) comprise the first main result of our investigation.

All of these pieces of the convergence depend on cosmological analogues of the PPN parameters. Noting the dependences, we expect three-point statistics of the convergence field to generically be functions of the parameters appearing in $\kappa^{(1)}$ and $\kappa^{(2)}$ above.  As an example, consider one of the simplest statistics, the skewness of the convergence field \cite{Refregier:2003xe}, $S_3 = {\left\langle\kappa^3\right\rangle}/{\left\langle\kappa^2\right\rangle^2}$, where the angle brackets denote averages over the survey area.  To lowest order, the denominator  depends on the first order convergence given by Eq.~(\ref{kappa1}), and numerator depends on a combination of the first and second order pieces: $\left\langle \kappa^3 \right\rangle = 3\left\langle \kappa^{(1)}\kappa^{(1)}\kappa^{(2)} \right\rangle +$~higher-order terms. Hence, the 2PN lensing deflection, including the cosmological analogues of the PPN parameters $\beta$ and $\varepsilon$, contributes to the three-point function of the convergence.

\section{Theories of gravity}

The indications of new gravitational phenomena are manifest in weak lensing through the cosmological analogues of the PPN parameters $\gamma$ at linear order, and $\beta$ and $\varepsilon$ at second order. We have shown that lensing depends in particular on the combination $(6-4\beta+4\gamma+3\varepsilon-2\gamma^2)$. It turns out that, for scalar-tensor theories of gravity at least, the combination $3 \varepsilon - 4 \beta$ depends simply upon $\gamma$. That $\varepsilon$ does not introduce a new degree of freedom has previously been noted \cite{Damour:1995kt}. We will show that the dependence of $\varepsilon$ upon $\beta$ is precisely cancelled when the sum of metric coefficients $g_{00} + g_{xx}$ required for lensing is taken. This means $\gamma$, or its cosmic analogue occasionally referred to as ``gravitational slip," is all that is required to model weak gravitational lensing to second order. This applies to scalar-tensor theories, including specific theories such as the well-studied Jordan-Brans-Dicke (e.g. Ref.~\cite{Will:1993ns}), $f(R)$ \cite{Carroll:2003wy}, and chameleon or ``symmetron"  \cite{Hinterbichler:2011ca} theories.

Consider a scalar-tensor theory of gravity with the action
\begin{equation}
S = \int d^4 x \sqrt{-g}  \left[ \frac{1}{16 \pi}\left( \phi R - \frac{{\vartheta}(\phi)}{\phi}(\partial\phi)^2 + 2 \phi \lambda(\phi)\right) - {\cal L}_{m}\right]~.
\label{eqn:genaction}
\end{equation}
Jordan-Brans-Dicke theory corresponds to the case $\lambda=0$. The $f(R)$ theory corresponds to setting $\vartheta=0$, $\lambda =\frac{1}{2}R (d\ln R/d\ln f - 1)$, and $\phi = f_{,R}$. The symmetron model
corresponds to a Higgs-like potential for $V(\phi)$ where $ \lambda(\phi) = -\phi V(\phi)$. All theories yield the same dependence of $\varepsilon$ upon $\beta$ and $\gamma$. We now focus on the $f(R)$ theory.

\subsection{f(R) Theory and Field Equations}

Consider the action
\begin{equation}
S = \int d^4 x \sqrt{-g}  \left[ \frac{c^2}{16 \pi} f(R) - {\cal L}_{m}\right]~,
\end{equation}
for which the field equations are
\begin{eqnarray}
F R_{\mu\nu} &=& \frac{8 \pi}{c^2}\left(T_{\mu\nu} - \frac{1}{2} g_{\mu\nu} T\right) + \left(F_{;\mu\nu} + \frac{1}{2}g_{\mu\nu}\Box F\right) - \frac{1}{2}\left(f - R F\right)g_{\mu\nu} \cr\cr
\Box F &=& \frac{1}{3}\left(\frac{8 \pi}{c^2} T + 2 f - R F\right)
\label{eqn:Feqn}
\end{eqnarray}
where $F \equiv \partial f/\partial R$. The second equation, equivalent to the scalar field equation of motion, guides the evolution of a new degree of freedom of the gravitational field as the Ricci scalar curvature ``comes to life'' as a dynamical field. Since the 2PN expansion in this theory has not been widely explored, we will provide some details of the calculations.

We consider a spacetime filled with non-relativistic matter in order to model the case of a static, spherically-symmetric mass $M$. Hence, the stress-energy tensor is
\begin{equation}
T_{\mu\nu} =  {\rho}\,  u_\mu u_\nu~.
\end{equation}
The fluid four-velocity satisfies $u^2=-1$ in the full spacetime. We define the ``conserved mass density'' $\rho^* \equiv \rho \sqrt{-g} u^t$; see Ref.~\cite{Will:1993ns}, Eq.~(4.77). By ignoring the cosmic fluid, consistency of the unperturbed field equations requires $f(R=0)=0$. A full treatment including the cosmic fluid has been carried out elsewhere \cite{Xie:2007gq} (see also \cite{Clifton:2008jq, Berry:2011pb, Lubini:2011pc}).
Since our focus is on scales within the Hubble horizon, as argued in the previous section, then for clarity we can simplify our calculations and use a static spacetime metric. Note that we have also carried out an analysis in a Robertson-Walker background, which agrees with the simpler results presented below when we restrict our attention to length scales within the Hubble horizon.

\subsection{Metric to order ${\cal O}$(${1/c^2}$)}
\label{sec:orderc2}

We now write the spacetime metric to order ${\cal O}$(${1/c^2})$ in the form
\be
ds^2 = -\left(1-\frac{1}{c^2}A_1\right) c^2 dt^2 + \left(1 + \frac{1}{c^2}B_1\right)d\vec x^2~.
\ee
Hereafter the subscripts indicate the expansion order in inverse-powers of $c^2$. The field equations give
\ba
\nabla^2(A_1+B_1) &=& -\frac{16 \pi}{f'(0)} \rho^* \cr\cr
\nabla^2 \nabla^2(A_1-2 B_1) - \frac{f'(0)}{3 f''(0)} \nabla^2(A_1-2 B_1) &=& -\frac{8 \pi}{3 f''(0)} \rho^{*}
\ea
where the Ricci scalar curvature to this order is $R_1 = \nabla^2(A_1-2B_1)$. The exterior solutions for $A_1$ and $B_1$ are no longer simply proportional to $1/r$ outside the mass source; the loss of Birkhoff's theorem means that we must take care to match the exterior solution onto the interior solution. See Ref.~\cite{Chiba:2006jp} for a detailed exposition of this theory to order ${\cal O}$(${1/c^2}$). To be definite, let us consider a constant-density source of radius $r_0$ where $\rho^{*} = 3M/(4 \pi r_0^3)$. Furthermore, let us define $m^2 \equiv f'(0)/(3 f''(0))$. Now we can show that for $r>r_0$
\ba
A_1 + B_1 &=& \frac{4 M}{f'(0) r}
\label{eqn:NpV}\\
A_1  - B_1 &=& \frac{4 M}{f'(0) r_0} i_1(m r_0) k_0(m r)
\label{eqn:NmV}
\ea
where $i_\ell(x)$ and $k_\ell(x)$ are modified spherical Bessel functions of the first and second kind. Solving for the desired potentials, we find
\ba
A_1 &=& \frac{4 M}{f'(0) r} \left[1 + \frac{r}{r_0} i_1(m r_0) k_0(m r) \right]
\label{eqn:Neqn}\\
B_1 &=& \frac{4 M}{f'(0) r} \left[1 - \frac{r}{r_0} i_1(m r_0) k_0(m r) \right]~.
\label{eqn:Veqn}
\ea
So if we equate $A_1=2 U$ and $B_1 = \gamma A_1$ then we infer
\ba
G^{-1} &=& \frac{1}{2}(1+\gamma) f'(0) \cr\cr
\gamma &=& 1 - \frac{2  \frac{r}{r_0} i_1(m r_0) k_0(m r)}{1+ \frac{r}{r_0} i_1(m r_0) k_0(m r) }~.
\label{frgamma}
\ea
In the limiting case $m r \gg 1$, $k_0(m r)\to 0$ and we recover GR with $\gamma=1$. In the case that $m r,\, m r_0 \ll 1$, we find $\gamma = 1/2$. Moreover, $\gamma$ grows monotonically from $1/2$ to $1$ as $r$ grows from $r \ll m^{-1}$ to $r \gg m^{-1}$. This highlights a feature of $f(R)$ gravity, that $1/2 \le \gamma \le 1$. We also learn that the contribution to the curvature does not vanish outside the massive body: it decays like a Yukawa potential with mass $m$. This spatial dependence of $\gamma$ manifests itself as small-scale fluctuations which are below the resolution of the particular weak lensing observations of interest here.

We also notice that the sum of metric potentials that appear in the lensing equations such as Eq.~(\ref{eqn:1PNlens}), i.e.~$A_1+B_1= 2(1+\gamma)U$ to order ${\cal O}(1/c^2)$ as given in Eq.~(\ref{eqn:NpV}), decays like $r^{-1}$ as in GR. Over the range of length scales for which these equations are valid, this means gravitational lensing is insensitive to the Yukawa potential -- lensing does not feel a fifth force. On the other hand, estimates of the mass $M$ based on orbital dynamics rely on $A_1$ and do feel the Yukawa potential in Eq.~(\ref{eqn:Neqn}).

\subsection{Metric to order ${\cal O}$(${1/c^3}$)}

To order ${\cal O}$(${1/c^3})$, the spacetime metric gains a term:
\be
ds^2 = -\left(1-\frac{1}{c^2}A_1\right) c^2 dt^2 + \left(1 + \frac{1}{c^2}B_1\right)d\vec x^2 +  \frac{2}{c^3}V_i c dt \, dx^i~.
\ee
The off-diagonal, $t-i$ component of the field equations yields
\be
\nabla^2V_i =\frac{16 \pi}{f'(0)} \rho^* v_i
\ee
where $v_i$ is the leading contribution to the spatial component of the source 4-velocity, $u_i = v_i/c + {\cal O}(1/c^3)$. The solution is $V_i = -4 \mathcal{V}_i/f'(0)$, where we define $\mathcal{V}_i$ implicitly as the solution to $\nabla^2\mathcal{V}_i = -4 \pi \rho^* v_i$. Substituting for $f'(0)$, we find the familiar result that $V_i = -2(1+\gamma)G \mathcal{V}_i$. The appearance of $\gamma$ is somewhat deceiving, however, since it is cancelled out by $G$, cf.~Eq.~(\ref{frgamma}). Consequently, the vector potential is also insensitive to the Yukawa potential in $A_1$.

\subsection{Metric to order ${\cal O}$(${1/c^4}$)}

As we proceed to higher order, we will ignore the time-dependence of the metric potentials and also ignore the contribution of the source velocities. What this means is that our results up to order ${\cal O}$(${1/c^3}$) are generally valid for a non-relativistic source, but higher order results are only valid for a static source.

We write the spacetime metric  to order ${\cal O}$(${1/c^4})$ as
\be
ds^2 = -\left(1-\frac{1}{c^2}A_1 - \frac{1}{c^4}A_2\right) c^2 dt^2 + \frac{2}{c^3}V_i c dt \, dx^i + \left(1 + \frac{1}{c^2}B_1+ \frac{1}{c^4}B_2\right)d\vec x^2~.
\ee
For our purposes it is sufficient to examine only the time-time ($\mu=\nu=0$) component of the perturbed field equations, whereupon
\be
\nabla^2(A_2+B_2) = -\frac{1}{4}  \nabla^2\left(A_1^2 +A_1 B_1 - \frac{3}{2} B_1^2 +\frac{1}{m^2}A_1 R_1\right)
+\frac{1}{6 m^2} R_1^2 + \frac{2 \pi}{f'(0)}\rho^*\left(A_1+2 B_1 + \frac{4}{3 m^2} R_1\right)~.
\label{eqn:lqeqn}
\ee
Defining  $\Phi_2$ implicitly as the solution to $\nabla^2\Phi_2 = -4 \pi \rho^* U$, then the solution to Eq.~(\ref{eqn:lqeqn}) is
\ba
A_2+B_2 &=& -\frac{1}{4}\left(A_1^2 +A_1 B_1 - \frac{3}{2} B_1^2 +\frac{1}{m^2}A_1 R_1\right) + (\beta' + \varepsilon')\Phi_2 \cr\cr
 (\beta' + \varepsilon')\Phi_2 &=& -\frac{1}{f'(0)} \int \frac{d^3 x'}{|\vec x - \vec x'|} \left[ \rho^* \left(\frac{1}{2}A_1+B_1 + \frac{2}{3m^2} R_1 \right) + \frac{f'(0)}{24\pi m^2} R_1^2\right]~.
\ea
Again, this is the sum of metric potentials we require for gravitational lensing. Since $R_1/m^2 = \frac{3}{2}(1-\gamma)A_1$, we find $A_2+B_2 = -\frac{1}{2}(5 -\gamma - 3 \gamma^2)U^2 + (\beta' + \varepsilon')\Phi_2$. Comparing with the general metric Eq.~(\ref{eqn:genmetric}), we find
\be
3 \varepsilon - 4 \beta = -5 +\gamma + 3 \gamma^2~.
\label{eqn:usefuleqn}
\ee
This combination, $3\varepsilon - 4 \beta$, is what appears in the lensing convergence, Eq.~(\ref{kappa2}).  Hence, in the context of this theory of gravity, the 2PN lensing terms depend solely on $\gamma$.

Phenomena other than lensing depend upon $\beta$ and $\varepsilon$ separately, so we turn to the spatial components of the gravitational field equations to find
\be
\nabla^2 B_2 + \frac{1}{3 m^2}\nabla^2 \nabla^2(A_2- 2 B_2) = \frac{2 \pi}{f'(0)} \rho^* B_1 + \frac{3}{8} \nabla^2(B_1^2)
-\frac{f'''(0)}{f'(0)}\nabla^2(R_1^2) + \frac{1}{12 m^2} C_2
\label{eqn:Qeqn}
\ee
where
\ba
&&C_2 \equiv
- \nabla^2\nabla^2(A_1^2 + A_1B_1+3B_1^2)+ \nabla^2(A_1\nabla^2 B_1 - 2 A_1 \nabla^2 A_1
+ 4 B_1 \nabla^2 A_1 - 8 B_1 \nabla^2 B_1) \cr\cr
&&\qquad\quad + (\nabla^2A_1 \nabla^2 B_1 - \nabla^2 A_1 \nabla^2 A_1 +  2 \nabla^2 B_1 \nabla^2 B_1)~.
\ea
In the GR limit, taking $m \to \infty$ and $f'''(0)\to 0$, the first term on the left-hand side and the first two terms on the right-hand side of Eq.~(\ref{eqn:Qeqn}) dominate. As a result, $B_2 = \frac{3}{2}U^2 - \Phi_2$ and $A_2 = -2 U^2 - 2 \Phi_2$, meaning that $\beta = \varepsilon=1$, as expected. In the limit $m \to 0$, and for $r^2 f''(0) \gg f'''(0)$, then the 2nd term on the left-hand side and the fourth term on the right-hand side of Eq.~(\ref{eqn:Qeqn}) dominate, leading to $\beta=1$ and $\varepsilon = 1/12$. The combination $3 \varepsilon - 4 \beta$ declines monotonically from $-1$ for GR down to $-15/4$ for $\gamma=1/2$.

\subsection{Other theories of gravity}

The $f(R)$ theory predicts $1/2 \le \gamma \le 1$. However, let us briefly consider a gravitational theory described by the action Eq.~(\ref{eqn:genaction}) with a potential $\lambda(\phi)$ and kinetic coupling $\theta(\phi)$. We presume that $\lambda$, $\theta$, and $\phi$ can be expanded perturbatively around their values on the background $\phi=\phi_0$, as described in explicit detail in Ref.~{\cite{Xie:2007gq}}. Next, following the same procedure as carried out in Sec.~\ref{sec:orderc2}, the equations to order ${\cal O}(1/c^2)$ are
\ba
\nabla^2(A_1+B_1) &=& - \frac{16 \pi}{\phi_0}\rho^* \cr\cr
\nabla^2(A_1-B_1) - m^2 (A_1-B_1) &=& - \frac{16 \pi}{\phi_0}\frac{1}{3 + 2 \theta(\phi_0)}\rho^*
\label{eq:bdpoisson}
\ea
where $m^2 =-2 \phi_0^2 \lambda''(\phi_0)/[3 + 2 \theta(\phi_0)]$. Considering a constant-density sphere of radius $r_0$ in the limit $r_0\to 0$, the solution for $\gamma$ is
\be
\gamma = 1 - \frac{2}{1 + e^{m r}[3 + 2 \theta(\phi_0)]}~.
\ee
We assume that $\phi_0 \lambda(\phi_0)$, $\phi_0^2 \lambda'(\phi_0)$, and the cosmic energy density are all negligible in comparison to the matter source $\rho$. Consequently, the only way to achieve a large value of $\gamma$ somewhere while still satisfying the Solar System constraint $\gamma = 1 + (2.1 \pm 2.3)\times 10^{-5}$ \cite{Bertotti:2003rm,Shapiro:2004zz} is if $-2 < \theta(\phi_0) < -3/2$ and $m r \gg 1$ on the scale of the Solar System. But even in such a case, at larger radii beyond the Solar System, though smaller than the Hubble scale, $\gamma$ asymptotes even closer to the GR value.

We can again see, in Eq.~({\ref{eq:bdpoisson}), that the lensing potential is insensitive to the Yukawa potential or fifth force, and instead decays like $1/r$ as in GR. What lensing does feel is the conformal factor $\phi_0$ that modulates the strength of the source in the above Poisson-like equation. Again, if that conformal factor can be determined elsewhere, such as by measurements of dynamics at radii within the fifth-force cutoff range, then lensing can give an undistorted view of the mass despite possible hindrance by a possible fifth-force. 

To extend our study, we may consider the chameleon mechanism, whereby the non-linear solution to the $\Box F$ equation (\ref{eqn:Feqn}) yields agreement with GR in high density regions such as the Solar System and Milky Way, but departure from GR leading to cosmic acceleration in lower density environments on larger scales. The requisite equations are little different from what we have already derived, but necessitate that we keep certain background terms that we have ignored in the preceding analysis. However we can borrow the results of Ref.~\cite{Hu:2007nk} (hereafter HS) to illustrate our points. To translate between the different notations, $A_1 = 2 {\cal A}_{HS}- 2{\cal B}_{HS}$, $B_1 = 2 {\cal A}_{HS}$, and $f =({R} + f_{HS})$. For a Universe containing only non-relativistic matter, the sum of potentials that appear in the lensing equation satisfy the equation
\be
\nabla^2\left(A_1 + B_1\right) = -\frac{16 \pi}{f'({R})}\rho  + \frac{1}{2}\Delta~,
\ee
whereas the individual potentials satisfy
\ba
\nabla^2 A_1 &=& -\frac{32 \pi}{3f'({R})}\rho  + \frac{1}{3}{R} + \frac{1}{3}\Delta\\
\nabla^2 B_1 &=& -\frac{16 \pi}{3f'({R})}\rho  -  \frac{1}{3}{R} + \frac{1}{6}\Delta
\ea
where $\Delta \equiv {R} - f({R})/f'({R})$. Here ${R}$ is the solution obtained from solving Eq.~(\ref{eqn:Feqn}) in the environment of stars, galaxies, and clusters as described by the non-relativistic mass density $\rho$. In order to satisfy Solar System constraints, $f \simeq {R}/G$ is required to lock the curvature to the density. To borrow the language of HS, this criterion helps define a critical value of curvature, above which GR is recovered. In such regions, $\Delta \ll {R},\, 8 \pi \rho/f'({R})$. In viable models of $f(R)$ gravity, then  the proportionality between the lensing potentials and the mass distribution is simple $1/f'(R)$. In regions where $\gamma = 1/2$, leading to $A_1 = 2 B_1$, then we can see that $\Delta,\, {R} \ll 8 \pi \rho/f'({R})$. This means the sum of the potentials is still generated by the density $\rho$ and not the novel curvature via ${R},\,\Delta$. Furthermore, if the bulk of the mass is located in a region in which $\gamma=1$, then the gravitational constant serving as the constant of proportionality between the lensing potential and mass is the same as that safely measured on smaller scales. 

The relationship between $\beta$ and $\varepsilon$ given in Eq.~(\ref{eqn:usefuleqn}) is valid for scalar-tensor theories with a single scalar field, as has been shown in Ref.~\cite{Damour:1995kt}. For a single scalar, there is a single new free function whose effect on the metric potentials is completely described by a single parametric function at the order ${\cal O}(1/c^2)$, namely $\gamma$. In the presence of additional scalars that couple non-minimally to gravity (e.g.~\cite{deRham:2010tw}), new functions may be necessary. It appears to be an open question whether vector-tensor and massive gravity theories,  such as the galileon \cite{Nicolis:2008in}, will also require additional functions to describe the departure from GR. Interesting signatures of galileon theories have recently been found for cluster lensing \cite{Wyman:2011mp}.

\section{Discussion}

We have explored what happens to the weak lensing distortion field at second order when the nonlinear behavior of gravity is altered. If we parameterize such deviations in a cosmological adaptation of the PPN framework, i.e.~starting from the metric (\ref{eqn:genmetric}), then we find corrections to the nonlinear weak lensing convergence that depend on the PPN nonlinearity parameters $\beta$ and $\varepsilon$. However, for a wide class of scalar-tensor gravity theories, we find that this combination generally cancels, thereby leaving the nonlinear correction to only depend on the ``gravitational slip'' $\gamma$. 
We have argued that the time-evolution of the PPN-like parameters leads to a negligible correction on scales well inside the Hubble horizon. 
We further derive the spatial dependence of $\gamma$, which is expected to be important only on sub-galactic scales and therefore 
would not impact the weak lensing observables discussed here.

Our results can be cast in terms of an effective $\Delta f_{\rm NL}$. Returning to the metric (\ref{eqn:fnl1}), then the sum of the ${\cal O}(1/c^4)$ potentials as appears in the lensing deflection is $-2\beta U^2 + \frac{3}{2}\varepsilon U^2 = -2(1-\Delta  f_{\rm NL})U^2 + 2(\frac{3}{4}+ \Delta f_{\rm NL})U^2$. Matching this sum with $A_2+B_2$ as computed in Section IIID, we determine that the additional non-Gaussianity introduced by gravity beyond GR is
\be
\Delta f_{\rm NL} = \frac{1}{8}(1 - 4\beta + 3 \varepsilon) = \frac{1}{8}(3\gamma^2+\gamma-4)~.
\ee
For the range $\frac{1}{2} \le \gamma \le 1$, we find $|\Delta f_{\rm NL}| < 1$. The tightest constraints to date come from the CMB \cite{Komatsu:2010fb}: $f_{\rm NL}=32 \pm 21$. Based on the scalings of the various contributions to the deflection angle given in Sec.~\ref{sec:oom}, a scalar-tensor theory would have to predict $\gamma \gg 10^2$ for these effects to be observable. But such a large value of $\gamma$ is not possible, and indeed this parameter has been constrained to be close to the GR value (e.g. \cite{Daniel:2010ky}). If this theory satisfies Solar System tests of GR then the non-Gaussianity of gravitational lensing relative to GR on scales below the Hubble scale is too weak to observe. However, it is worth noting that modifications to nonlinear gravitational dynamics can further impact $f_{\rm NL}$ via modifications to nonlinear structure formation. 
Overall, our results suggest that looking for the signs of new gravitational physics in the linear regime, in the evolution of large-scale clustering over cosmic time scales, e.g. as a consequence of ``gravitational slip",  remains a valid strategy.

If the Solar System tests are satisfied, then the gravitational constant measured in the Solar System is $G = 1/f'$. This is the same gravitational coupling as appears in gravitational lensing: for a mass $M$, the deflection relies on the sum of potentials $A_1 + B_1 = 4 M/f' r$. This means that sub-horizon gravitational lensing may provide a clean, direct method to measure mass in spite of a potential fifth-force resulting from a scalar-tensor modification of gravity.

\begin{acknowledgments}

We thank Wayne Hu, Mark Wyman, Scott Dodelson, and Christopher Berry for useful conversations. This research was carried out in part at the Jet Propulsion Laboratory, run by the California Institute of Technology under a contract from NASA, and Dartmouth College and was funded through the JPL Strategic University Research Partnership (SURP) Program.  We also acknowledge the support of the Kavli Institute for Cosmological Physics at the University of Chicago through grants NSF PHY-0114422 and NSF PHY-0551142 and an endowment from the Kavli Foundation and its founder Fred Kavli.

\end{acknowledgments}


\end{document}